\documentclass[preprint,12pt]{elsarticle}

\usepackage{amssymb}
\usepackage{url}
\usepackage{longtable}
\usepackage{color}

\definecolor{redmar}{rgb}{.8,0,0}

\newcounter{bla}

\journal{Computer Physics Communications}
\newcommand{\ta}[1]{#1\hspace{-.5em}/\hspace{.20em}} 

\begin{document}
\sloppy

\begin{frontmatter}



\title{EKHARA 3.0: an update of the EKHARA Monte Carlo event generator}


\author[a,b]{Henryk Czy\.z \corref{author}}
\author[a]{Patrycja Kisza}

\cortext[author] {Corresponding author.\\\textit{E-mail address:} henryk.czyz@us.edu.pl}
\address[a]{Institute of Physics, University of Silesia,
PL-41500 Chorz\'ow, Poland}
\address[b]{Helmholtz-Institut, 55128 Mainz, Germany}

\begin{abstract}
 The Monte Carlo event generator EKHARA was upgraded during last years.
 The upgrades presented here contain: a) the inclusion of new final 
 states $e^+e^-\to e^+e^-\eta $,  $e^+e^-\to e^+e^-\eta'$, $e^+e^-\to e^+e^- \chi_{c_i} $ and $e^+e^-\to e^+e^- \chi_{c_i} (\to J/\psi (\to \mu^+\mu^-)\gamma)$;
  b) new $\gamma^*-\gamma^*-P$ transition form factors and c) the radiative 
  corrections to the reactions $e^+e^-\to e^+e^-P $. For the upgrades
  a) and b), we present here only the pieces missing in other publications,
   mostly algorithms used in the phase space generation. The radiative corrections
  are presented here for the first time. A new algorithm of the phase
  space generation for the reaction $e^+e^-\to e^+e^-P \gamma$ being its
  main part. A comparison with GGRESRC generator is presented. Big
  differences between the radiative corrections calculated by the EKHARA and
   the GGRESRC generators
   are observed.

\end{abstract}


\end{frontmatter}



{\bf NEW VERSION PROGRAM SUMMARY}

\begin{small}
\noindent
{\em Program Title: EKHARA 3.0}                                          \\
{\em Licensing provisions: GPLv3}                                   \\
{\em Programming language: FORTRAN 77}\\                                   
{\em Supplementary material: The following publications are relevant for the
  upgrades presented in Section \ref{two}: [3,4,5,6] }
  \\
{\em Journal reference of previous version: [1] }                  \\
{\em Does the new version supersede the previous version?:} Yes  \\
{\em Reasons for the new version:} Major upgrades.\\
{\em Summary of revisions:}  The upgrades contain the inclusion of the new final 
 states $e^+e^-\to e^+e^-\eta $,  $e^+e^-\to e^+e^-\eta'$, $e^+e^-\to e^+e^- \chi_{c_i} $ and $e^+e^-\to e^+e^- \chi_{c_i} (\to J/\psi (\to \mu^+\mu^-)\gamma)$,
   new $\gamma^*-\gamma^*-P$ transition form factors and the radiative 
  corrections to the reactions $e^+e^-\to e^+e^-P $. \\
{\em Nature of problem:}\\
  The program is constructed to help in 
  measurements of transition form factors at meson factories.
  To do this one needs a good description
   of the already existing data and calculation of the radiative 
  corrections.  \\
{\em Solution method:}
   The models of various form factors were developed in 
     [3,4,5,6] and are
   implemented in the program. For the radiative corrections a new
   algorithm of generation of the phase space for the reactions
   $e^+e^-\to e^+e^-P \gamma$ was developed and is presented here.\\
{\em Additional comments including Restrictions and Unusual features:}\\
  The program needs a quadruple precision version of a FORTRAN compiler.
   \\

\end{small}

\section{Introduction}
\label{int}
The hadron physics and especially hadron-photon interactions entered
precision era \cite{Actis:2010gg} some years ago. The push towards
precision was mostly caused by the disagreement of the measured
 \cite{Bennett:2006fi} and
 calculated within the Standard Model \cite{Keshavarzi:2018mgv,Hagiwara:2017zod,Jegerlehner:2017gek,Davier:2017zfy}
 muon anomalous magnetic moment $(g-2)_\mu$. This might be a hint of a signal
 from physics beyond the Standard Model. With the new measurement
 of the muon anomalous magnetic moment under way \cite{Gioiosa:2017oda},
  the current  4 $\sigma$
 disagreement might become the first laboratory observation of the
 physics beyond the Standard Model. An effort to reach the precision
 expected in the new muon anomalous magnetic moment measurement
  in its calculation has to be made, and in fact it has already started
 as a ``g-2 Theory Initiative'' \cite{TI}. It is  a common venture
 of theorists from lattice and phenomenology communities and experimental
 communities involved in measurements of hadronic cross sections and
 transition form factors. The knowledge of the pseudoscalar transition form factors
 with a good accuracy, in the range of the kinematic invariants, which is
 as wide as possible, is a prerequisite for improving the accuracy of the
 light-by-light contributions. The error on this part is now at the same level
 as the a error on the hadronic vacuum polarisation contributions to the $(g-2)_\mu$
 and thus its reduction is as important as the error reduction on the hadronic
  vacuum polarisation contribution.
 
The increasing requirements for precision in the measurements of the transition
form factors (amplitudes) $\gamma^*-\gamma^*-{\rm hadrons} $, rise 
a question of the precision of the Monte Carlo generators used in the experimental
analyses.  In the latest measurements
 \cite{Aubert:2009mc,BABAR:2011ad,Uehara:2012ag} of the most important
 for the evaluation of the light-by-light contributions to the muon anomalous magnetic
  moment, pseudoscalar transition form factors,
  Monte Carlo generators based on structure function approach were used
  \cite{TREBSBST,Druzhinin:2014sba}. The accuracy of the structure function
  approach is very much dependent on the event selection used
  \cite{Rodrigo:2001kf} and if possible should be cross checked with exact
  results. A step towards this goal was done in this paper, where the most relevant
   radiative corrections at NLO were implemented into the event generator EKHARA.
  
   The paper is organised in the following way: in Section \ref{two} the algorithms
   used in the previous upgrades of the EKHARA code, which were not covered
   in the previously published papers, are described. In Section \ref{three}
   the implementation of the NLO radiative corrections and the tests
    of the code are presented in details.
    In Section \ref{four} the size of the radiative corrections for event selections
    close to the experimental ones is discussed and comparisons with 
  the GGRESRC generator \cite{Druzhinin:2014sba} are shown.
  In Section \ref{five} an overview of the EKHARA software structure and
   users guide are given. Conclusions are drawn in Section \ref{six}.
  
\section{Upgrades from version 2.0 to 2.3}
\label{two}
The mode $e^+e^-\to e^+e^-\pi^+\pi^-$ was not changed.

 The $\gamma^*-\gamma^*-P$ transition form factors were 
 upgraded twice (release 2.1 and 2.3).
 In the first upgrade the models presented in \cite{Czyz:2012nq}
 were implemented. In the second upgrade the form factors
  coming from the models developed in \cite{Czyz:2017veo} were added. 
  The two models developed in \cite{Czyz:2017veo}
  should be used as a default, as they describe the largest class 
   of experimental data. The other models can be used in tests
   of the model dependence of various entities (experimental efficiencies etc.).
   The simulation of the phase space for the reaction $e^+e^-\to e^+e^-P$
   is identical, up to the bug fixed in
  the azimuthal angles generation, to the one used in version 2.0
  \cite{Czyz:2010sp} (see also \cite{Schuler:1997ex}, from where
  the algorithm used in \cite{Czyz:2010sp} was adopted).
    Due to the bug, only one half of the allowed azimuthal angular range
   was covered in  version 2.0. This bug was affecting only simulations
   with cuts imposed on azimuthal angles as the relative
   angles between the momenta and the polar angles were correct.

  In the release 2.2 the model of the $\gamma^*-\gamma^*-\chi_{c_i}$ and  
   $\gamma^*-J/\psi^*-\chi_{c_i}$ form factors
  developed in \cite{Czyz:2016xvc,Czyz:2016lwq} was implemented 
  to allow a simulation of the reactions 
 $e^+e^-\to e^+e^- \chi_{c_i} $ and 
 $e^+e^-\to e^+e^- \chi_{c_i} (\to J/\psi (\to \mu^+\mu^-)\gamma)$.
  The generation of the phase space in the reaction $e^+e^-\to e^+e^- \chi_{c_i} $
  is again identical to the one in \cite{Czyz:2010sp}. 

   For the reaction 
  $e^+e^-\to e^+e^- \chi_{c_i} (\to J/\psi (\to \mu^+\mu^-)\gamma)$
 $(i=0,1,2)$
  we write

   \begin{eqnarray}
     d\sigma(e^+(p_1)e^-(p_2)\to e^+(q_1)e^-(q_2)\mu^+(q_4)\mu^-(q_3)\gamma(k))
   &=& \nonumber \\
  &&\kern-140pt|M_i|^2 dLips_5(p_1+p_2;q_1,q_2,q_3,q_4,k)\, .
  \label{chi}
  \end{eqnarray}

  The matrix element $M_i$ is described in \cite{Czyz:2016lwq}.
   Here we report
  the details of the phase space parameterisation, which allowed 
  for absorption of the peaking behaviour of the phase space.
   We write the five-particle phase space in the following form

   \begin{eqnarray}
 dLips_5(p_1+p_2;q_1,q_2,q_3,q_4,k) &=& \nonumber \\
    && \kern-200pt dLips_3(p_1+p_2;q_1,q_2,Q_\chi)
   \frac{dQ^2_\chi}{2\pi} dLips_2(Q_\chi;Q_\psi,k)
   \frac{dQ^2_\psi}{2\pi} dLips_2(Q_\psi;q_3,q_4)\, ,
   \label{pschi}
  \end{eqnarray}
 where $Q_\chi = q_3+q_4+k$ and $Q_\psi=q_3+q_4$. We generate at first 
  the invariant mass $Q_\chi^2$ of the virtual $\chi_{c_i}$ meson
   within limits $4m_\mu^2<Q_\chi^2< (\sqrt{s}-2m_e)^2$, unless
  the user specifies otherwise (see Section \ref{five}), with
   $s=(p_1+p_2)^2$ and $m_\mu(m_e)$  being muon (electron) mass respectively.
  To absorb
  the peak coming from the $\chi_{c_i} $ propagator the following change
  of variables was performed

    \begin{eqnarray}
    &&Q_\chi^2 = M_{\chi_{c_i}}\Gamma_{\chi_{c_i}}
         \tan\left(\frac{y}{M_{\chi_{c_i}}\Gamma_{\chi_{c_i}}}\right) 
   + M_{\chi_{c_i}}^2\ ,
     \ \ \ \ 
      y = y_{min} + \Delta y \cdot r \ ,   \nonumber  \\
      && \Delta y =  y_{max} - y_{min}, \ \ \ \ 0<r<1  \ ,   \nonumber  \\
      && y_{min(max)} = M_{\chi_{c_i}}\Gamma_{\chi_{c_i}} 
     \arctan\left(  \frac{Q_{\chi \ min(max)}^2 - M_{\chi_{c_i}}^2 }{M_{\chi_{c_i}}\Gamma_{\chi_{c_i}}}\right)   
     \, ,
   \label{BWchi}
  \end{eqnarray}
 where $M_{\chi_{c_i}}$ ($\Gamma_{\chi_{c_i}}$) is the mass (width) 
   of the $\chi_{c_i}$ meson.

  The same is done for the generation of the $Q_\psi^2$, which is generated
   as the second variable, with the change of the $\chi_{c_i} $ mass (width)
  to the $J/\psi$ mass (width):
   $M_{\chi_{c_i}}\to M_{J/\psi}$, $\Gamma_{\chi_{c_i}}\to \Gamma_{J/\psi}$. The limits
   read $4m_\mu^2< Q_\psi^2 < Q_\chi^2$, unless a user has required 
  more stringent cuts (see Section \ref{five}). The angles of the muons
  are generated flat in the rest frame of the $Q_\psi$ and than transformed
  to the $Q_\chi$ rest frame. The photon angles are generated flat
  in the $Q_\chi$ rest frame and later transformed to the initial $e^+e^-$
  center of mass frame together with the muon and anti-muon four-momenta.
   The $dLips_3(p_1+p_2;q_1,q_2,Q_\chi)$ is generated in the same way as for the
    reaction $e^+e^-\to e^+e^- \chi_{c_i} $ following the description
   in \cite{Czyz:2010sp}, where the $\chi_{c_i}$ mass was replaced by 
   the invariant mass $Q_\chi^2$.

   In the channel, where contributions coming from all three $\chi_i$
  intermediate states are included, we use three channel Monte Carlo
  method to absorb the three peaks coming from $\chi_i$ propagators.
   In each channel we use the change of variables described 
   above to absorb the peaks. The probability of using  
    a given channel is chosen according to tuned 'a priori' weights. 
   This simple method works efficiently because  the interferences
   between the amplitudes are negligible \cite{Czyz:2016lwq}. That results from
   the fact that the $\chi_i$ resonances are narrow and well separated.

\section{The radiative correction to the reaction $e^+e^-\to e^+e^-P $}
\label{three}
  \subsection{Virtual radiative corrections}
\begin{figure}[h]
\begin{center}
\includegraphics[width=3.5cm]{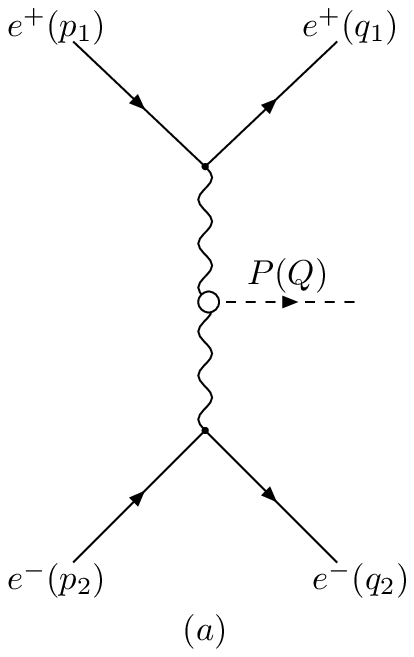}
\includegraphics[width=3.5cm]{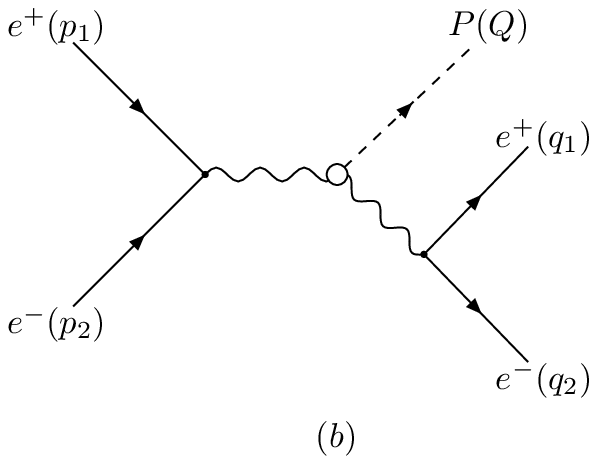}
\includegraphics[width=3.5cm]{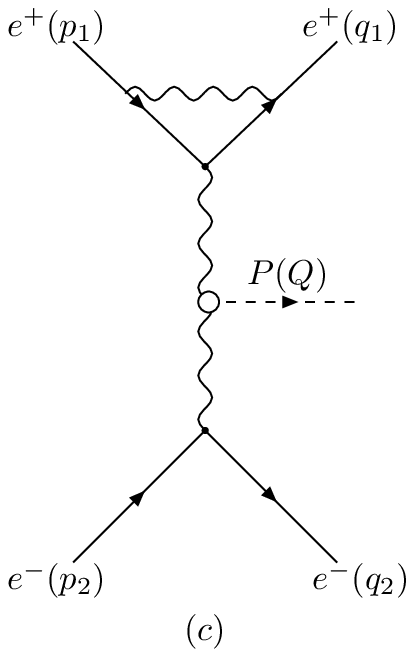}\\
\includegraphics[width=3.5cm]{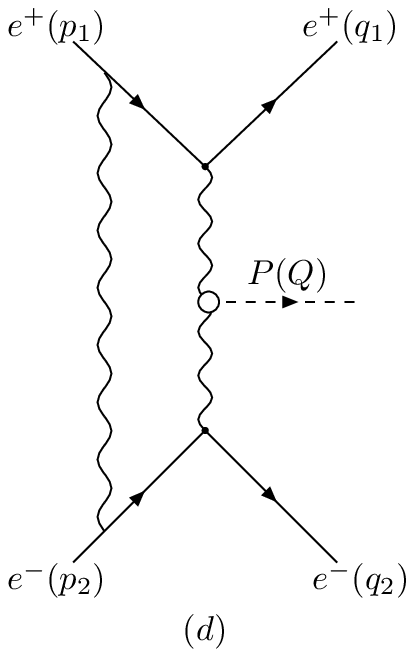}
\includegraphics[width=3.5cm]{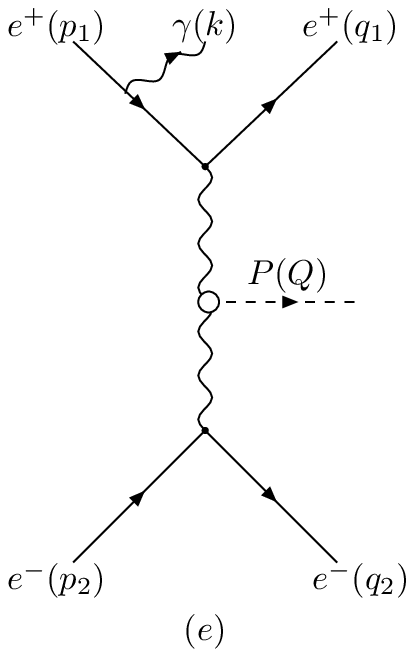}
\includegraphics[width=3.5cm]{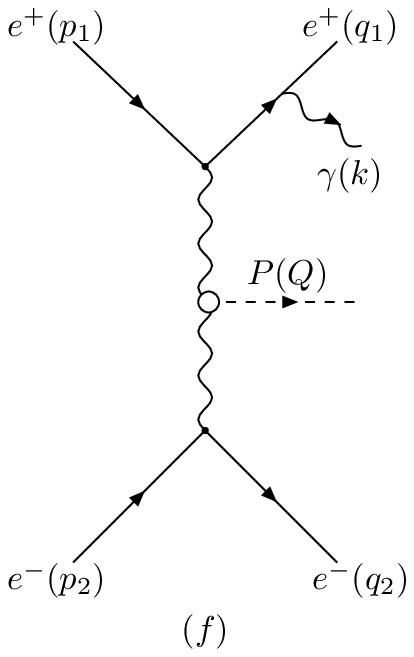}
\caption{ Representative sample of diagrams contributing to the amplitude $e^+e^-\to e^+e^-P(\gamma) $.
  \label{diag}}
\end{center}
\end{figure}

  The vertex virtual correction (Fig. \ref{diag}(c) and a similar diagram
   with corrections to the electron line) are known already for some time
  \cite{Barbieri:1972as} and we use in the code the expressions from that paper. The formulae
  were checked later by many groups (see for example \cite{Bonciani:2003ai}).
  The form of the corrections is
  \begin{equation}
   (LO): \bar v(p_1) \gamma^\mu v(q_1) \to
    (NLO):   \bar v(p_1) \left([1+F_1(t_1)]\gamma^\mu -
      \frac{F_2(t_1)}{4m_e}[\ta q,\gamma^\mu] \right)v(q_1) \, , \nonumber \\
\label{vertex}
 \end{equation} 
  with $q=q_1-p_1$ and $t_1=(q_1-p_1)^2$. The functions $F_1$ and $F_2$
   are given in Eqs. (2.19-2.10) of \cite{Barbieri:1972as}. The corrections
     coming from $F_2$ are negligible for all event selections shown in these paper.
  
   We have included into the code only radiative corrections
   to the t-channel diagram (Fig. \ref{diag}(a)) as the s-channel diagram
    (Fig. \ref{diag}(b))
  is important only for the configurations where both final leptons are observed.
  Moreover, the kinematic region, where both the  t- and s-channel contributions are
  of the similar size, is far from being reached by any experiment due to the small value
   of the cross section in this kinematical region.
   The contributions from five point functions (Fig. \ref{diag}(d)) were found to be
   negligible \cite{vanNeerven:1984ak} and are not considered here. Yet it is
    worthwhile to reconsider these corrections for configurations with two final
    leptons observed at large angles as they are model dependent. 
   We plan to include the remaining radiative corrections,
     in a separate publication
     \cite{CKS}, where also their model dependence will be studied.

     Following \cite{Barbieri:1972as} we use a fictitious photon mass to regulate
     the infrared singularities. It is also used in the real emission part,
     where in the phase space parameterisation a massive photon was
      assumed (see Section \ref{real} for details).
   
  The vacuum polarisation corrections are included in a fully factorised
   and resummed form

     \begin{eqnarray}
       &&\kern-30ptM_{LO+virt}(e^+e^-\to e^+e^-P) \to  M_{LO+virt}\cdot\frac{1}{1-\Delta\alpha(t_1)}
       \cdot  \frac{1}{1-\Delta\alpha(t_2)}\, ,\nonumber \\
       &&\kern-30ptM_{e+f}(e^+e^-\to e^+e^-P\gamma) \to  M_{e+f}(e^+e^-\to e^+e^-P\gamma)
        \cdot\frac{1}{1-\Delta\alpha(t_1')}
       \cdot  \frac{1}{1-\Delta\alpha(t_2)}\nonumber \, , \\ \label{vacpol}
  \end{eqnarray}
     with $t_2 = (p_2-q_2)^2$, $t_1' = (p_1-q_1-k)^2$. Analogously one adds
     the radiative corrections to the diagrams with a photon emitted from the
      electron lines.
     The $\Delta\alpha(t)$ is taken from \cite{alphaQED}
      (see also \cite{Jegerlehner:2017zsb,Jegerlehner:2011mw}).
     
  \subsection{Real radiative corrections}\label{real}

   The matrix element describing the reaction
\begin{equation}
 e^+(p_1)e^-(p_2)\to e^+(q_1)e^-(q_2)P(Q)\gamma(k)
 \label{notation}
 \end{equation} 
 was calculated using Feynman diagrams shown in Fig. \ref{diag} (e) and
  (f) and similar diagrams, where photon is emitted from the electron
  line.
  As stated already, the five point functions (Fig. \ref{diag} (d)) were found negligible
    \cite{vanNeerven:1984ak}. They cancel  the infrared
    singularities from the interference
   between the diagrams with the photon emitted from the positron lines
   and with the  photon emitted from the electron lines. 
     Thus that interference has to be neglected for consistency.

  It is convenient to parameterise the phase space of the
  reaction Eq.(\ref{notation}) in the following way \cite{book}

     \begin{eqnarray}
 &&\kern-30pt\int dLips_4(p_1+p_2;q_1,q_2,Q,k) = \nonumber \\
 &&\kern-15pt \frac{1}{(2\pi)^8}
   \frac{1}{4\sqrt{\lambda(s,m_e^2,m_e^2)}}
  \int\displaylimits_{(m_P+m_e+m_\gamma)^2}^{(\sqrt{s}-m_e)^2}dM_3^2
   \int\displaylimits_{t_3^-}^{t_3^+}dt_3  
  \int\displaylimits_{0}^{2\pi}d\phi_2
   \nonumber \\
    &&  
  \kern-15pt 
  \frac{1}{4\sqrt{\lambda(M_3^2,t_3,m_e^2)}}
  \int\displaylimits_{(m_e+m_\gamma)^2}^{(M_3-m_P)^2}dM_2^2
   \int\displaylimits_{t_2^-}^{t_2^+}dt_2
   \int\displaylimits_{0}^{2\pi}d\phi_P 
   \frac{1}{4\sqrt{\lambda(M_2^2,t_2,m_\gamma^2)}}
    \int\displaylimits_{t_1^-}^{t_1^+}dt_1
   \int\displaylimits_{0}^{2\pi}d\phi_1 \, ,\nonumber \\  
   \label{phspacenlo}
  \end{eqnarray}
with
     \begin{eqnarray}
  &&\lambda(a,b,c) = a^2+b^2+c^2-2ab-2ac-2bc\, ,\   \ \  t_1 = (p_1-k)^2\, ,
  \nonumber \\ 
  &&M_3^2 = (Q+k+q_1)^2 = (p_1+p_2-q_2)^2\, ,
  \   \ \ \ \ \ \ t_3 = (p_2-q_2)^2 \, ,\nonumber \\ 
  && M_2^2 = (k+q_1)^2 = (p_1+p_2-q_2-Q)^2\, , \   \ \ \ \ \ \
  t_2 = (p_2-q_2+Q)^2\, ,
  \nonumber \\ 
   \label{phspacenlodef}
  \end{eqnarray}
     $m_P,m_e$ and $m_\gamma$ being pseudoscalar, electron and
     a fictitious photon mass, respectively, 
and
     \begin{eqnarray}
  &&t_3^{\pm} = m_e^2+M_3^2-\frac{1}{2s}\left\{
   s(s+M_3^2-m_e^2)\mp 
   \sqrt{\lambda(s,m_e^2,m_e^2)\lambda(s,M_3^2,m_e^2)} \right\}\, ,
  \nonumber \\ 
  &&t_2^{\pm} = m_e^2+M_2^2-\frac{1}{2M_3^2}\Bigg\{
   (M_3^2+m_e^2-t_3)(M_3^2+M_2^2-m_P^2) \nonumber \\ 
  && \kern+160pt\mp 
   \sqrt{\lambda(M_3^2,m_e^2,t_3)\lambda(M_3^2,M_2^2,m_P^2)} \Bigg\}\, ,  
  \nonumber \\ 
  &&t_1^{\pm} = m_e^2+m_\gamma^2-\frac{1}{2M_2^2}\Bigg\{
   (M_2^2+m_e^2-t_2)(M_2^2+m_\gamma^2-m_e^2) \nonumber \\ 
  && \kern+160pt\mp 
   \sqrt{\lambda(M_2^2,m_e^2,t_2)\lambda(M_2^2,m_\gamma^2,m_e^2)} \Bigg\}\, .  
  \nonumber \\ 
   \label{phspacenlodef1}
  \end{eqnarray}

 In this way all the peaks appearing in the matrix element can be easily 
   absorbed for the contributions from Fig. \ref{diag} (e) and
  (f). 
 To do that we use the following changes of variables
     \begin{eqnarray}
  t_i = -e^{-z_i} , i=2,3;  \ \ \ M_2^2 = m_e^2 + e^{y}; \ \ \
   t_1 = m_e^2 - e^{-z_1}\, .
   \label{varchange}
  \end{eqnarray}

     In the above formulae, $e^a$ should be read as $1$ GeV$^2 \cdot e^a$.
       For simplicity, the unit $1$ GeV$^2$ was dropped in all formulae. 
     
   The first two changes of variables absorb peaks coming from
  the virtual photons propagators, the third (fourth) one the peak
  coming from the positron  propagator in the diagram with photon
   emitted from final (initial) positron line. The last two 
   changes of variables have
  to be present simultaneously as the leading contribution comes
   from the interference of the diagram (e) and the diagram (f). 
 
   The user introduced cuts on $t_3$
   are used to alter the generation limits.
   Other user cuts are just rejecting events generated
   outside the allowed phase space.
   The cuts on $t_3^{min}<t_3<t_3^{max}$ change the maximal
   allowed value of $M_3$ ( $M_3^{max}$) if both  $t_3^{min}$ and $t_3^{max}$
   are bigger (lower) than $t_3^c = 2m_e^2-\sqrt{s}m_e$. It reads
   \begin{eqnarray}
    && M_{3,max}^2 = \frac{2m_e^4+t_3^{min}s+\sqrt{\Delta(t_3^{min})}}{2m_e^2}
     \ , \ \ {\rm for} \ \ t_3^{min}>t_3^c, t_3^{max}>t_3^c \nonumber \\
      && M_{3,max}^2 = \frac{2m_e^4+t_3^{max}s+\sqrt{\Delta(t_3^{max})}}{2m_e^2}
     \ , \ \ {\rm for} \ \ t_3^{min}<t_3^c, t_3^{max}<t_3^c
     \label{limits1}
  \end{eqnarray}
   with $\Delta(t) = t s (t-4m_e^2)(s-4m_e^2)$.

   As the matrix element contains also the contributions coming
    from the photons
   emitted from the electron line we use two-channel Monte Carlo, where
   in the second channel the parameterisation of the phase space is identical
   to the one described above with the change $p_2 \leftrightarrow p_1$
   and $q_2 \leftrightarrow q_1$. Within that two-channel scheme the
   phase space parameterisation is written as

         \begin{eqnarray}
          \int dLips_4(p_1+p_2;q_1,q_2,Q,k) =
           \int\displaylimits_{0}^{1}dr_0
           \left[\theta\left(\frac{1}{2}-r_0\right)C_1
             +\theta\left(r_0-\frac{1}{2}\right)C_2\right] \nonumber \\
    \label{mctwochannel}
  \end{eqnarray}
         with $\theta$  being a Heaviside step function and $C_i,\ \ i=1,2$
         the parameterisations of the phase space in channels $1$ and $2$.
         The parameterisation in the channel  $1$ reads
         
         \begin{eqnarray}
   &&\kern-15pt C_1 = \frac{1}{(2\pi)^5}
   \frac{\Delta M_3^2}{2\sqrt{\lambda(s,m_e^2,m_e^2)}}
  \int\displaylimits_{0}^{1}dr_1
   \int\displaylimits_{0}^{1}dr_2  
  \frac{ \Delta z_3}{4\sqrt{\lambda(M_3^2,t_3,m_e^2)}}
   \nonumber \\
    &&  
  \kern-15pt 
  \cdot \int\displaylimits_{0}^{1}
      dr_3
   \int\displaylimits_{0}^{1} dr_4 
  \frac{\Delta y \Delta z_2}{4\sqrt{\lambda(M_2^2,t_2,m_\gamma^2)}}
    \int\displaylimits_{0}^{1}dr_5 \frac{ \Delta z_1}{f_1+f_2}
   \int\displaylimits_{0}^{1}dr_6
    \int\displaylimits_{0}^{1}dr_7 
     \int\displaylimits_{0}^{1}dr_8 \, ,\nonumber \\  
      \label{mctwochannel1}
         \end{eqnarray}
     where   
      
     \begin{eqnarray}
      && M_3^2 = M_{3,min}^2 + \Delta M_3^2 \cdot r_1 , \ \
       \Delta M_3^2 =M_{3,max}^2-M_{3,min}^2\, ,\nonumber \\
       &&   z_3 = z_3^{min} + \Delta z_3 \cdot r_2 , \ \
       z_2 = z_2^{min} + \Delta z_2 \cdot r_4, \ \ z_1 = z_1^{min} + \Delta z_1 \cdot r_5
       \, ,\nonumber \\
       &&z_i^{min} = -\log(-t_i^{min}), \ \
       \Delta z_i = -\log\left(\frac{t_i^{max}}{t_i^{min}}\right), \ \ i=2,3 \nonumber \\
               &&   z_1^{min} = -\log(-\tilde t_1^{min}), \ \
       \Delta z_1= -\log\left(\frac{\tilde t_1^{max}}{\tilde t_1^{min}}\right)\, ,\nonumber \\
       && \tilde t_1^{max(min)} = m_\gamma^2\nonumber \\
       &&-\frac{(M_2^2+m_e^2-t_2)(M_2^2+m_\gamma^2-m_e^2)+(-)
         \lambda^{1/2}(M_2^2,m_e^2,t_2)  \lambda^{1/2}(M_2^2,m_e^2,m_\gamma^2)}
             {2M_2^2}
             \, ,\nonumber \\
       && y = y_{min} + \Delta y  \cdot r_3, \ \  y_{min} = \log(m_\gamma(2m_e+m_\gamma))
                   \, ,\nonumber \\
      && \Delta y = \log\left(\frac{(M_3-m_P)^2-m_e^2}{m_\gamma(2m_e+m_\gamma)}\right),
              t_1 = \tilde t_1 + m_e^2\, ,\nonumber \\
    &&  \kern-15pt   f_1 =
       \frac{-1}{t_3(M_2^2-m_e^2)t_2(t_1-m_e^2)}, \ \ 
            \phi_2 = 2\pi \cdot r_{6}, \ \  \phi_P = 2\pi \cdot r_{7}, \ \  
          \phi_1 = 2\pi \cdot r_{8}  \, .\nonumber \\
       \label{mctwochannel1a}
         \end{eqnarray}
 The function $f_2$ is obtained with $f_1$ with  the change $p_2 \leftrightarrow p_1$
 and $q_2 \leftrightarrow q_1$. The parameterisation
 of the phase space in the second channel ($C_2$) is obtained
  from $C_1$ with the same substitutions.

  From the generated variables described above one can calculate the four-momenta
  of all final particles. Again, we give here only formulae for the channel 1
  as the channel 2 is obtained in the same way with the substitutions
   $p_2 \leftrightarrow p_1$
  and $q_2 \leftrightarrow q_1$. Moreover, as it is possible to write some
  of the expressions  given below in two or more analytically equivalent forms,
  we give here only the ones used in the code. They were chosen to obtain
   formulae which are numerically stable.

  The azimuthal angle of the final electron ($\phi_2$) is generated in the
  initial $e^+e^-$ center of mass frame with positron momentum along the
  z-axis: $p_1 = (\sqrt{s}/2,0,0,p)$, $p=\sqrt{s/4-m_e^2}$. This frame is called
    the LAB frame from now on.
   The energy ($E_2$), the length of the momentum ($lq_2$)
   and the cosine of the polar angle ($\theta_2$) of the final electron
   can be calculated,
     in the same frame, from the generated invariants
         \begin{eqnarray}
           &&   \kern-30pt  E_2 = \frac{s-M_3^2+m_e^2}{2\sqrt{s}}, \ \ 
           lq_2  = \frac{\lambda^{1/2}(M_3^2,s,m_e^2)}{2\sqrt{s}} , \ \
           \cos(\theta_2) = \frac{M_3^2-s-2t_3 +3m_e^2}{4\cdot p\cdot lq_2}
           \, .\nonumber \\
          \label{costh2}
         \end{eqnarray}
         The azimuthal angle of the pseudoscalar ($\phi_P$) is generated in the rest
         frame of the four-vector $p_1+p_2-q_2$, where the z-axis is pointing
         the initial positron momentum $p_1 = (\tilde E_1,0,0,\tilde p_1)$. Here
         $\tilde E_1 = \frac{M_3^2+m_e^2-t_3}{2M_3}$. In this frame
         $p_2-q_2 = (\tilde E_2,0,0,-\tilde p_1)$, with
         $\tilde E_2 = \frac{M_3^2-m_e^2+t_3}{2M_3}$. In the code,
        for
         numerical stability reasons,  the expression
          $\tilde p_1 = \sqrt{\tilde E_2^2-t_3}$ is used to calculate $\tilde p_1$.
           In this frame, the pseudoscalar energy
         ($E_P$), the length of the pseudoscalar momentum ($lq_P$) and
         the cosine of the pseudoscalar polar angle ($\theta_P$) are
         given by
               \begin{eqnarray}
           &&   \kern-30pt  E_P = \frac{M_3^2-M_2^2+m_P^2}{2M_3}, \ \ 
           lq_P  = \sqrt{E_P^2-m_P^2} ,  \nonumber \\
        &&   \kern-30pt   \cos(\theta_P) = \frac{t_3-t_2+m_P^2-
             \frac{(M_3^2-m_e^2+t_3)(M_3^2-M_2^2+m_P^2)}{2M_3^2}}
                {2\cdot \tilde p_1\cdot lq_P}
           \, .
          \label{costhP}
         \end{eqnarray}
              After being calculated, the pseudoscalar four vector is transformed
               into the LAB frame.
               
          The azimuthal angle of the final positron ($\phi_1$) is generated in the rest
         frame of the four-vector $p_1+p_2-q_2-Q$, where the z-axis is pointing
         the initial positron momentum $p_1 = ( E_1^*,0,0,p_1^*)$. Here
         $E_1^* = \frac{M_2^2+m_e^2-t_2}{2M_2}$. In this frame
         $p_2-q_2-Q = (E_2^*,0,0,-p_1^*)$ with
         $E_2^* = \frac{e^y+t_3}{2M_3}$. In the code the expression
         $p_1^* = \sqrt{E_2^{*,2}-t_2}$ is used to calculate $p_1^*$.
         The $p_1+p_2-q_2-Q$ rest frame is also the $q_1+k$ rest frame,
          thus the final positron and the final photon momenta differ only by  a sign. 
          In this frame, the final positron energy
         ($E_1$), the length of its momentum ($lq_1$),
          the cosine of its polar angle ($\theta_1$) and
            the photon energy ($E_\gamma$) are
         given by
               \begin{eqnarray}
           &&   \kern-30pt  E_1 = \frac{M_2^2-m_\gamma^2+m_e^2}{2M_2}, \ \ 
                 lq_1  = \frac{\sqrt{(e^y+m_\gamma^2-2m_\gamma M_2)
                      (e^y+m_\gamma^2+2m_\gamma M_2)}}{2M_2} ,  \nonumber \\
                 &&   \kern-30pt   E_\gamma = \frac{e^y+m_\gamma^2}{2M_2}, \ \
                  \cos(\theta_1) = \frac{m_e^2+m_\gamma^2-
             \frac{(e^y+m_\gamma^2)(M_2^2+m_e^2-t_2)}{2M_2^2}}
                {2\cdot  p_1^*\cdot lq_1}
           \, .
          \label{costh1}
         \end{eqnarray}
               From the rest frame of the $p_1+p_2-q_2-Q$ four-momentum
                to the LAB frame
               the four vectors are transformed in two steps. First to
               the $p_1+p_2-q_2$ rest frame and than to the LAB frame.
               In this way the same subroutine can be used for both
                transformations. It consists of a boost and three elementary rotations.
                It was checked numerically that, after transforming all four vectors
                 to the LAB frame, $q_1+q_2+Q+k = p_1+p_2$ within 28-digits accuracy.

     \subsection{Tests of the code}

     The code is using in its bulk part the quadruple numerical precision,
     with exceptions described in Section \ref{five}. All the tests described below
     were performed with a precision of one half of a per mile or better.
     We cover here only the tests of the newly developed part. The tests
     of the previously developed parts of the code are covered in
      \cite{Czyz:2006dm,Czyz:2010sp,Czyz:2012nq,Czyz:2016lwq}.
     
     The matrix element of the LO contribution to
     the cross section of the reaction $e^+e^-\to e^+e^-P$
     was tested in \cite{Czyz:2010sp}, thus one does not have to test the part
     of the virtual radiative corrections $\sim F_1$ as they are proportional
     to the same matrix element. The part of the virtual radiative corrections
     proportional to $F_2$ was calculated, using trace method to sum over polarisations,
     independently by two of the authors.
     As this contribution is negligible, no further tests were performed.

     For the matrix element describing the reaction $e^+e^-\to e^+e^-P\gamma$
     two independent codes were constructed. One using helicity amplitude method,
     where sum over helicities was done numerically, and one using trace method
     to sum over polarisations. For the calculations using trace method
     the symbolic manipulation system FORM \cite{Kuipers:2012rf} was used
     and a FORTRAN code was produces based on its output. Even if
       the code uses quadruple precision, the code constructed using the trace
      method is not numerically stable around kinematical points with few
      invariants appearing in the denominators of the expression being close to zero
      simultaneously. As there are almost no numerical cancellations in
      the formulae, which use the helicity amplitude method, these formulae
      are free from such problems. An agreement up to 28 digits was found
      between the results obtained with the two described methods for
      all the phase space points with the exception of the situations described
      above. The formula, which uses the helicity amplitude method is used
      in the distributed version of the code.

      The phase space parameterisation, together with the change of variables
      described in  Section \ref{real}, was tested comparing the phase space
      volume calculated within that parameterisation and the volume calculated
      with an independent code, which uses a flat cascade-like parameterisation
      \cite{book}. A very good agreement was found for all tested energies,
      in the range $1\ GeV< \sqrt{s} < 11 \ GeV$, and for physical
        masses of the pseudoscalar
        particles ($\pi^0,\eta, \eta'$).

        The differential cross sections, when one sums the contributions
        with and without a real photon emission should not depend on
        the fictitious photon mass ($m_\gamma$) introduced as a regulator. We use
        a parameter $\lambda$ ($m_\gamma = \lambda m_e$) to set its
        size in the code. The recommended value for $\lambda$ is 0.01.
        For $\lambda\sim 0.1$ some of the differential cross sections
        start to depend on this parameter with deviations bigger than
        the one set as a goal for technical accuracy in this code (0.05\%).
        For $\lambda=0.001$ and $\lambda=0.0001$ the differential
        cross sections were identical to the one obtained with
        $\lambda=0.01$ within the errors of about 0.05\%. Due to this
        small cut-off, the infrared divergent part in the virtual corrections,
        which is negative, is bigger than 1 resulting in the negative cross
        section. As a result, one cannot generate unweighted event sample
         and only weighted events can be used.
      
\section{The size of the radiative corrections and comparisons with GGRESRC Monte Carlo generator}
\label{four}

\begin{figure}[h]
\includegraphics[width=13.cm]{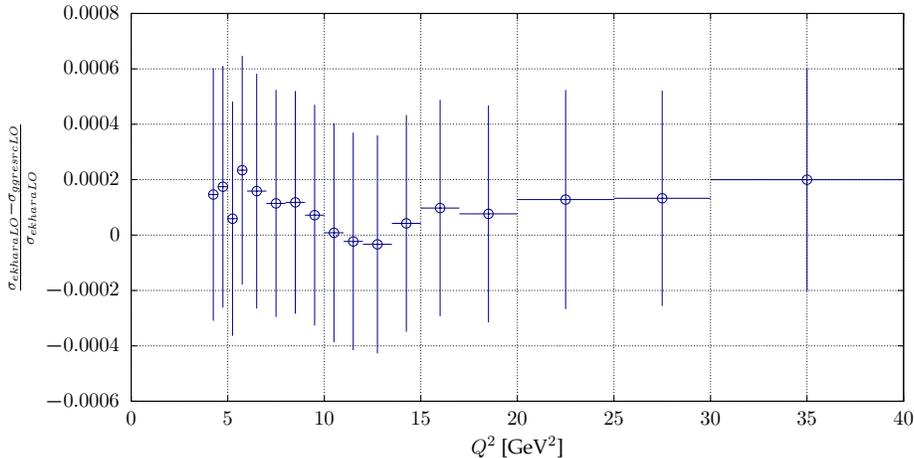}
\caption{Comparison of the EKHARA and GGRESRC generators at LO; 
 \ \ \ \ \ \ \ \  \ \ \  \  \ \ \ \ \ \ \ \  \ \ \  \  $Q^2=-(p_1-q_1)^2$, $\sqrt{s} = 10.58 {\rm GeV}$.
\label{LOcomp}}
\end{figure}

There exists a Monte Carlo event generator \cite{Druzhinin:2014sba}, GGRESRC,
were the radiative corrections to the  reaction $e^+e^-\to e^+e^-P$
were included using a structure function method. This generator was used
in the BaBar analysis to measure the $\gamma-\gamma^*-P$
transition form factors \cite{Aubert:2009mc,BABAR:2011ad}. The accuracy
of the structure function method depends a lot on the event selection
 (see for example \cite{Rodrigo:2001kf}).
It is thus important to check it against exact calculations, whenever
 possible. 
 For simulation of the reaction $e^+e^-\to e^+e^-P$,
at LO level the EKHARA and GGRESRC Monte Carlo generators are
very similar. When the same form factor is used in both codes
we have observed an agreement at a level of 0.04\%. 
 This was already observed in  \cite{Druzhinin:2014sba} for
 $e^+e^-\to e^+e^-\pi^0$. We have checked that also for $\eta$ and $\eta'$
 integrated cross sections, angular and energy distributions of all final
 particles are identical for both generators in a single tag mode. To obtain
 this agreement a VMD transition form factor used in GGRESRC was implemented
 in the EKHARA Monte Carlo generator. In all the comparisons between the generators
 shown in this paper that form factor is used. An example of these comparisons
  is shown
 in Fig. \ref{LOcomp}. We have restricted the invariant
 $-0.18\ {\rm GeV^2}< (p_2-q_2)^2<0\ {\rm GeV^2}$ and calculated the cross section
 in bins of  $Q^2=-(p_1-q_1)^2$ as shown in Fig. \ref{LOcomp}. The bins coincide
 with the bins used by BaBar collaboration \cite{Aubert:2009mc}.

  At NLO it is not a straightforward task to compare the two event generators
  as in the formulae used in  GGRESRC generator the 'final' photon is integrated
  out. The final photon would come from the  the diagram Fig. \ref{diag} (f). Yet
  the interference between the amplitudes coming from  Fig. \ref{diag} (e)
  and Fig. \ref{diag} (f) gives substantial contributions to the matrix element
  squared and the identification 'final' or 'initial' photon is not possible.
  One could of course define the final or initial photon on the bases of
  being closer to the initial or final lepton, but then one would need to integrate
  the photons which are closer to the final lepton, to be close to the formulae
  used in GGRESRC generator. There is one more difference between the generators:
  in GGRESRC the corrections to the line of the untagged lepton are not included.
  We thus start with the check how big they are in the EKHARA generator to disentangle
  these two different effects. We do it for the event selection close to the one used by BaBar.
  We use here kinematical variables (angles, energies) in the center of mass frame
   of the initial leptons.
  We require that the final positron and pion polar angles are in the range 
  $20^\circ < \theta_{\pi,e} < 160^\circ$. We also put cuts on polar angle of $\pi^0 e^+$
  system $\cos(\theta_{e\pi})>0.99$ and  on a variable
   $r =(\sqrt{s}- E_{e\pi}-p_{e\pi})/\sqrt{s} <0.075$.

  \begin{figure}
\includegraphics[width=13.cm]{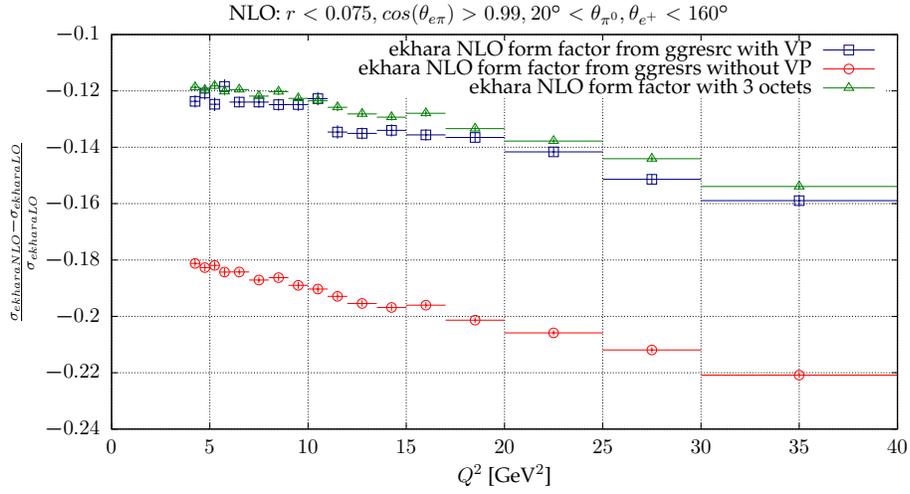}
\caption{ The size of the radiative corrections predicted with EKHARA Monte Carlo generator.
\label{NLOcomp}}
  \end{figure}
  
   \begin{figure}
\includegraphics[width=13.cm]{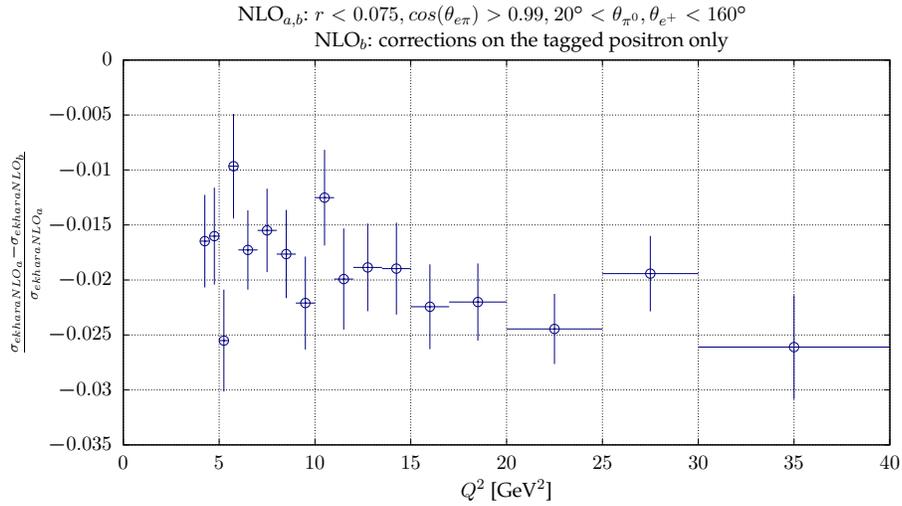}
\caption{ The complete corrections compared to the corrections only to the tagged
  lepton line.
\label{NLOcomp1}}
    \end{figure}
  
   The results are shown in Fig. \ref{NLOcomp}.
  The complete radiative corrections for that event selection are
   negative and amount from
   12 \% to 16\% depending on the range of the $Q^2$ invariant.
    The corrections coming from photon vacuum polarisation amount,
    for this event selection,
  to 6-7.5\% depending on the tagged invariant and are positive.
  The size of the corrections depends only slightly  on the form factor,
   as
  shown in Fig. \ref{NLOcomp}. The cross sections predicted with these two
  different form factors differ up to 35\%. Yet, the radiative corrections
   as a fraction of LO cross section differ at most by 1\%.
  The form factors used in this
  comparison were: the VDM form factor from GGRESRC generator and
  the form factor based on 3 octet model from \cite{Czyz:2017veo}.

    If we switch off
   the corrections to the untagged electron
   line we find out that indeed, as stated in  \cite{Druzhinin:2014sba},
    the dominant contribution comes from
   the tagged line. The difference, amounting to 1.5-2.5 \%, is shown
    in Fig. \ref{NLOcomp1}.

\begin{figure}
\includegraphics[width=13.cm]{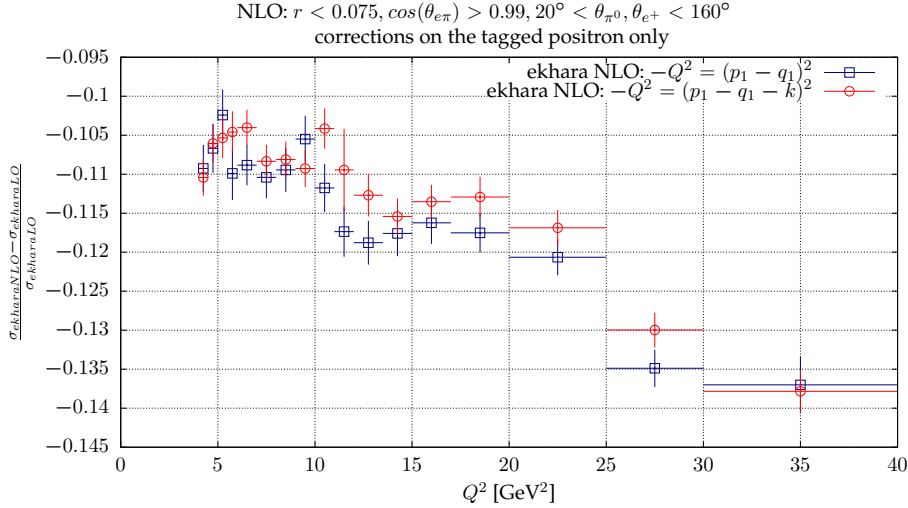}
\caption{ The size of radiative corrections compared for binning in variable
   $Q^2 = -(p_1-q_1-k)^2$ and $Q^2 = -(p_1-q_1)^2$.
\label{NLObinning}}
\end{figure}

One has to
mention here that the $Q^2 = - (p_1-q_2)^2$
  is not the invariant for which the form
   factor is calculated. The correct invariant reads
   $(p_1-q_1-k)^2$. The imposed cuts assure that the second invariant
   is close to zero. The size of the radiative corrections, if one uses the correct
   invariant, is also different.
    Yet, the difference is small,  as shown in Fig. \ref{NLObinning}. 
    
\begin{figure}
\includegraphics[width=13.cm]{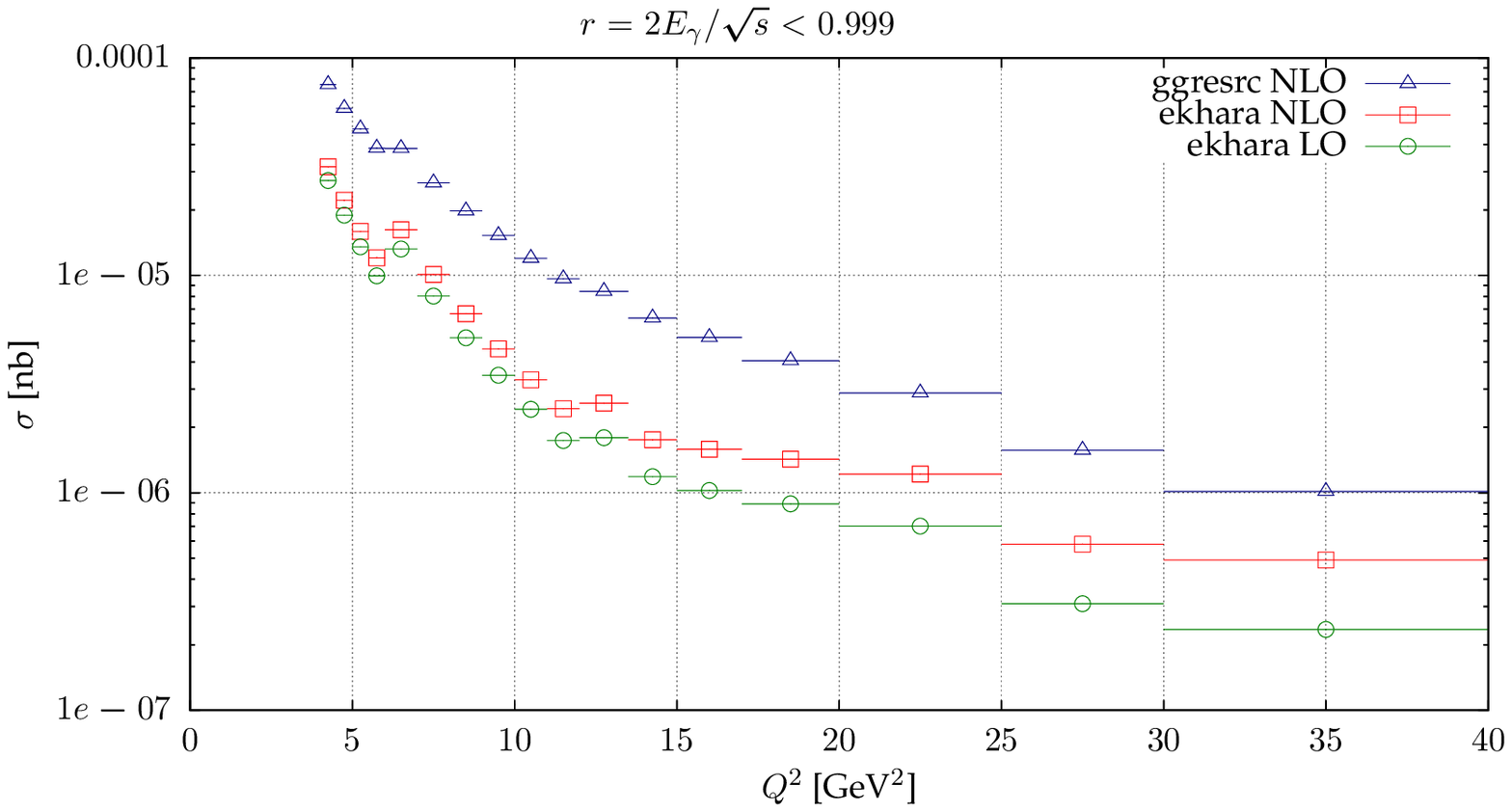}
\caption{ Comparison between EKHARA and GGRESRC Monte Carlo generators.
\label{ekhgg1}}
\end{figure}
      \begin{figure}
\includegraphics[width=13.cm]{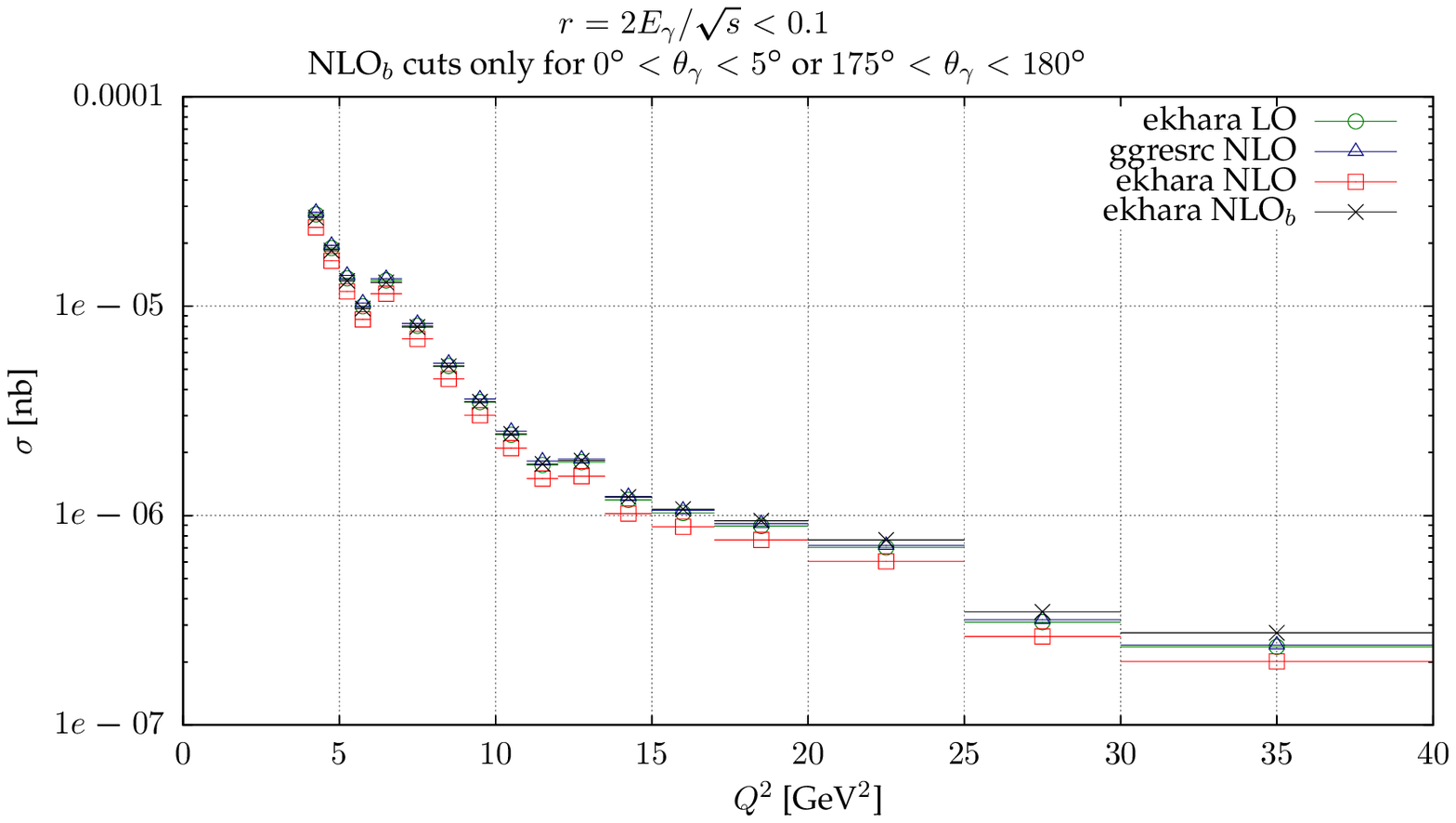}
\caption{ Comparison between EKHARA and GGRESRC Monte Carlo generators.
\label{ekhgg2}}
\end{figure}
     \begin{figure}
\includegraphics[width=13.cm]{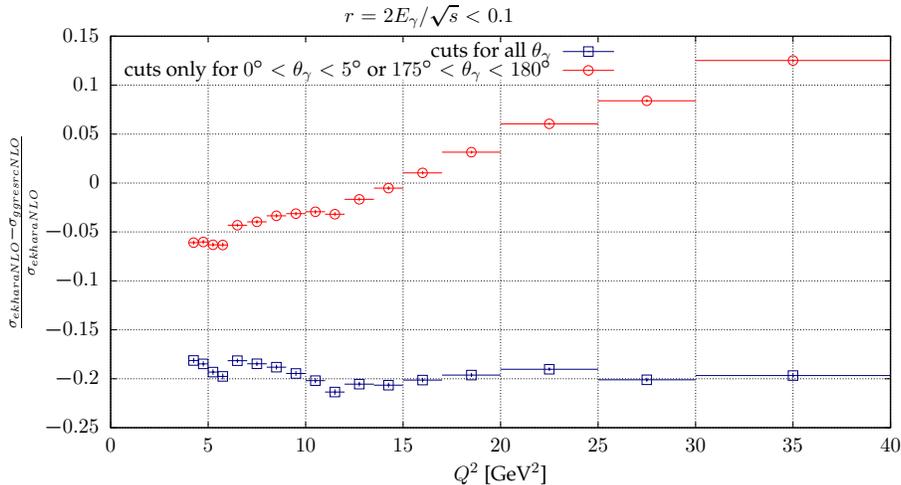}
\caption{ Comparison between EKHARA and GGRESRC Monte Carlo generators.
\label{ekhgg3}}
\end{figure}

As the TREBSBST \cite{TREBSBST} generator is not publicly available
in its version with radiative corrections, we compare here only our results
 with GGRESRC generator.
    To see the differences in the NLO predictions between the EKHARA and GGRESRC
    event generators, we show in Fig. \ref{ekhgg1}
     the results with no direct angular cuts and
    the untagged invariant in the range $-0.18 \ {\rm GeV^2} < (p_2-q_2)^2 <0$,
    as a function of the tagged invariant. We do not impose any cut on the emitted
    photon energy in EKHARA and use Rmax=0.999 in GGRESRC.
    This should correspond to the situation were one does not
     include any direct cut on the photon variables. Big differences
     are observed. We were not able to trace back the source of the difference.

 Smaller differences are observed (Fig. \ref{ekhgg2}), when one imposes a cut-off
     on the photon energy. Yet, as in GGRESRC the photons are partly integrated,
     the compared cross sections are not defined identically and one cannot
     expect a complete agreement. To come closer to the GGRESRC we have applied
     cuts on the photon energy only if the photon polar angles are within 5 degrees
     from the initial leptons direction. The results are also shown in Fig. \ref{ekhgg2}.
     As expected the results come closer, yet they are not in agreement.
       The relative difference is shown in Fig. \ref{ekhgg3}.

\section{The software structure and the users guide}
\label{five}

\subsection{An overview of the code structure}

The  overview of the code structure is given here and
 for completeness  we repeat in part the description given in \cite{Czyz:2010sp}. A more detailed users guide is a part of the distributed package.
 
Let us start with an overview of the directory structure of the distribution.
EKHARA is distributed as a source code.
The code of the Monte Carlo generator is located in the
directory {\tt ekhara-routines}. 
The main source file of EKHARA is {\tt ekhara.for}.
There are other source files in the directory {\tt ekhara-routines},
which are automatically included:
\begin{itemize}
 \item
the $e^+e^- \to e^+e^- P$ modes are
implemented in
     {\tt routines\_1pi.inc.for}             
and its supplementary histograming routines are given in
{\tt routines-histograms\_1pi.inc.for};
 \item
the $e^+e^- \to e^+e^- \pi^+\pi^-$  mode
is coded in 
{\tt routines\_2pi.inc.for},
its supplementary histograming routines are in
{\tt routines-histograms\_2pi.inc.for}
and helicity-amplitude routines are given in
{\tt routines-helicity-aux.inc.for};
  \item
the $e^+e^-\to e^+e^- \chi_{c_i} $ and $e^+e^-\to e^+e^- \chi_{c_i} (\to J/\psi (\to \mu^+\mu^-)\gamma)$ modes are
implemented in
     {\tt routines\_chi.inc.for}             
and its supplementary histograming routines are given in
{\tt routines-histograms\_chi.inc.for};
\item
the NLO corrections to $e^+e^- \to e^+e^- P$ modes are
implemented in
{\tt routines\_1pi\_1ph.inc.for}. It uses partly 
 routines from {\tt routines\_1pi.inc.for}.
 Its supplementary histograming routines are given in
{\tt routines-histograms\_1pi.inc.for};
\item
 the routines for the matrix and vector manipulations
 are located in {\tt routines-math.inc.for};
 \item
 in {\tt routines-user.inc.for} 
 several routines, which can be 
 changed by a user in order to customise the operation of EKHARA, are collected;
 they handle the data-card reading, 
 the reporting of events, the form factor evaluation, the filling of 
the histograms,
 the application of additional phase space cuts, etc.; 
 \item
 all common blocks are  included from the
 file {\tt common.ekhara.inc.for}.
This file contains the detailed comments 
on the explicit purpose 
of the most important common variables.
 \item
   the routines from the package alphaQEDc17 \cite{alphaQED} to calculate
   the vacuum polarisation corrections are contained in files:
   {\tt common.h,constants.f,constants\_qcd.f,dalhadshigh17.f,}\\
   {\tt dalhadslow17.f,
     dalhadt17.f,dggvapx.f,hadr5n17.f,leptons.f,}\\
    {\tt vacpol\_alphaQEDc17.inc.for}
\end{itemize}

The operation of the EKHARA generator requires the following steps:
\begin{enumerate}
  \item the initialisation,
  \item the event generation,
  \item the finalisation.
\end{enumerate}

The main directory of the distributed version contains a {\tt readme.txt} file
with a short description how to compile, run and test the program in the
regimes described above. 
It is suggested to use the {\tt Makefile}, which is placed
in the main directory.
An example of the full set of input files and the plotting environment is
supplied in the {\tt Env} sub-directory. 
If one uses the distributed {\tt Makefile}, the content of
the {\tt Env} sub-directory will be put into the {\tt EXE} sub-directory
together with an executable {\tt ekhara.exe}.

\subsection{the I/O scheme and files}
\label{subsec:io}

All the input files of EKHARA are supposed to
be located in the same directory as the main executable, {\tt ekhara.exe}.
There are the following types of the input files:
random seeds, the parameter input, data-cards and histogram settings.
An example of the full set of input files 
can be found in the {\tt Env} directory.

All the output files of EKHARA 
are written into {\tt ./output} sub-directory.
There are the following types of the output files:
logs of execution, histograms and events.

\subsubsection*{The input files}

The main input file is called {\tt input.dat}. It contains all global settings,
which are explained in this file as well.

The channel-dependent parameters are collected in ``data-cards''  {\tt card\_1pi.dat}, {\tt card\_2pi.dat}  and {\tt card\_chi.dat}.
These data-cards allow to set the total energy, types of included amplitudes 
and kinematic cuts. A detailed description can be found in comments
within these files.
In {\tt card\_1pi.dat} one can also use the {\tt piggFFsw} switch
in order to select the form of the two photon pseudoscalar
transition form factor. The recommended values are 9 or 10 as these are
the form factors which were fitted to the widest data set \cite{Czyz:2017veo},
 both in the
 space-like and time-like regions.

The channel-dependent histograming settings are given 
in the files   {\tt histo-settings\_1pi.dat}, {\tt histo-settings\_2pi.dat},
{\tt histo-settings\_chi.dat} and {\tt histo-settings\_1pi\_1ph.dat}. 

\subsubsection*{The output and the logging}
The main execution log file is  {\tt output/runflow.log}.
It contains  main information about the operation mode and status
of EKHARA, this information is also partly written into the standard
output (i.e., the console).
At the end of a successful execution, the total cross section
is reported to {\tt output/runflow.log} and also to 
the standard output.

A non-standard behaviour of the MC generator
is reported into {\tt output/warnings.log},
while the critical problems in the event generator operation
are reported into the file {\tt output/errors.log}.

In the case of a correct operation,  {\tt output/errors.log} 
and {\tt output/warnings.log} should remain empty.
We strongly recommend to keep track on this issue and report to the
authors any warnings or errors. In the NLO mode one can ignore
warnings about negative weights in the part without a photon.
Yet in this case only weighted events can be used.

\subsubsection*{The output: histograms and plotting scripts}

When histograming is allowed through settings in the 
{\tt input.dat}, the plain text files with
the histogram data are saved at the end of the generator
execution.

\begin{itemize}
 \item In the $e^+e^- \to e^+e^- \pi^+\pi^-$ mode the file
       {\tt histograms\_2pi.out} contains
       the data for $d \sigma / dQ^2$ histogram.
       One may use the plotting script {\tt doplots.sh}
       from directory {\tt histo-plotting\_2pi}
       in order to plot this histogram 
       (an installed {\tt Gnuplot} is required). 

 \item In the $e^+e^- \to e^+e^- P$ modes there is a wide set
       of histograms stored in the files
       {\tt histo<Number>.<variable>.dat},
       where {\tt<Number>} stands for the histogram number
       and {\tt <variable>} is the histograming variable acronym.
       
       One can use the plotting script {\tt do-everything.sh}
       in the directory {\tt histo-plotting\_1pi}
       in order to plot all the histograms and collect them
       into a single postscript file.
       An installed \LaTeX \ system is required for the latter. 
       
       One can use the plotting script {\tt doplots.sh}
       in the directory {\tt t1-t2-bars\_1pi}
       in order to plot the 3D-bar graph, which shows
       the event distribution in two variables: $t_1$ and $t_2$.

       In the NLO mode the files {\tt histo\_th\_electron.dat}, 
       {\tt histo\_th\_positron.dat} and {\tt histo\_th\_pseudoscalar.dat} 
       contain the data for $\Delta \sigma$ in the polar angles
        $\Delta\theta_{e^-}$,
       $\Delta\theta_{e^+}$ and  $\Delta\theta_{\pi^0, \eta, \eta^{\prime}}$
       histograms respectively. $\Delta \sigma$ is the integrated
        cross section in a given bin.
       
       One can use the plotting script {\tt do-everything.sh}
       in the directory {\tt histo-plotting\_1pi\_1ph}
       to plot all the histograms and collect them
       into a single postscript file.

     \item In the $e^+e^-\to e^+e^- \chi_{c_i} $
       and $e^+e^-\to e^+e^- \chi_{c_i} (\to J/\psi (\to \mu^+\mu^-)\gamma)$
       modes  the file {\tt histograms\_chi.out} contains
       the data for $d \sigma / dQ^2$ histogram.
       
       One can use the plotting script {\tt doplots.sh}
       in the directory {\tt histo-plotting\_chi} in order to plot 
       histogram and to obtain a single postscript file.
     
\end{itemize}
As the histograms are stored as plain text files
the user can use also her/his favourite plotting programs 
to visualise the histograms.

\subsubsection*{The output: events}

The generated four-momenta of the particles are stored in the following variables accessible through common blocks:
\noindent
\begin{longtable}{p{0.3\textwidth} p{0.7\textwidth}}
{\tt p1}&initial positron, 
\\
{\tt p2}&initial electron,
\\
{\tt q1}&final positron,
\\
{\tt q2}&final electron,
\\
{\tt qpion}&final pseudoscalar ($e^+e^- \to e^+e^- P(\gamma)$ modes),
\\
{\tt k\_hp}&final photon ($e^+e^- \to e^+e^- P \gamma $ modes),
\\
{\tt qu}&final $\chi_{c_i}$ ($e^+e^- \to e^+e^- \chi_{c_i}$ modes),
\\
{\tt q3}& final $\mu^-$ in  $e^+e^-\to e^+e^- \chi_{c_i} (\to J/\psi (\to \mu^+\mu^-)\gamma)$  modes),
\\
{\tt q4}& final $\mu^+$ in  $e^+e^-\to e^+e^- \chi_{c_i} (\to J/\psi (\to \mu^+\mu^-)\gamma)$  modes),
\\
{\tt k1}& final photon in  $e^+e^-\to e^+e^- \chi_{c_i} (\to J/\psi (\to \mu^+\mu^-)\gamma)$  modes),
\\
{\tt pi1, pi2}&final pseudoscalars ($e^+e^- \to e^+e^- \pi^+\pi^-$ mode).
\\
  {\tt   contribute}& { weights NLO.} 
\end{longtable}

    The weights in the NLO mode allow
    to calculate a cross section, given in nanobarns,
     for any event selection using
     a formula $\Delta\sigma_k = \frac{\sum_i w_i}{N_k}$, with
     $k=0,1$, where $k=0$ stands for events without a photon
     and $k=1$
     stands for events with one photon.
      In the output the {\tt k\_hp} is a zero four vector for events without a photon.
      $w_i$ is the weight,  $N_0$ is the number of events with no photons
      and $N_1$ is the number of events with one photon. The sum span
      over all events for a given event selection.
 
In the standalone regime we suggest to 
use the routine {\tt reportevent\_1pi} defined in the file 
{\tt routines-user.inc.for}, which is called
automatically for every accepted unweighted event
($e^+e^- \to e^+e^- P$ modes only). 
In the NLO mode the routine {\tt reportevent\_1pi\_1ph},
defined in this same file, reports every event used to calculate 
 the cross section from the weighted events. 
In the chi\_c modes the routine {\tt reportevent\_chi}, 
which can be found in the file {\tt routines-user.inc.for},
reports momenta for every accepted unweighted event.
In the distributed version this routine writes the
events to the file {\tt output/events.out}
 when {\tt WriteEvents} flag is on.

\subsection{Selected procedures}

The top-level interface to the Monte Carlo generator is provided 
by the routine
\noindent
\begin{longtable}{p{0.3\textwidth} p{0.7\textwidth}}
{\tt EKHARA(i) }& \parbox[t]{0.7\textwidth}{
               {\tt i = -1}: initialise, \\
               {\tt i = 0}: generate event(s), \\
               {\tt i = 1}: finalise.
               }
\end{longtable}
\noindent
Only this routine should be called from an external program,
when one uses EKHARA in the event-by-event regime.
An example is provided in {\tt ekhara-call-example.for}.

In order to  describe briefly the ``internal'' structure
of EKHARA, we list several important routines.

\noindent
\begin{longtable}{p{0.3\textwidth} p{0.7\textwidth}}

{\tt EKHARA\_INIT\_read}  & \parbox[t]{0.7\textwidth}{the reading the input files and datacards,}
\\
{\tt EKHARA\_INIT\_set}   & \parbox[t]{0.7\textwidth}{the initialisation of the MC loop and mappings,}
\\
{\tt EKHARA\_RUN }        & \parbox[t]{0.7\textwidth}{the MC loop execution,}
\\
{\tt EKHARA\_FIN}         & \parbox[t]{0.7\textwidth}{the MC
 finalisation and the saving the results.}
\end{longtable}

\subsection{Compilation instructions}
\label{sec:Install}

Being distributed as a source code the program
does not require installation, but a compilation and a linking are needed.
EKHARA does not need any specific external libraries, but requires
\begin{itemize}
 \item a FORTRAN 77 compiler which supports the quadruple precision,
 \item a C compiler.
\end{itemize}
The current version of the program was tested on the following platforms
 : Linux (Ubuntu 14.04, Ubuntu 16.04).
The program distribution contains the {\tt Makefile},
with targets:  {\tt default64} - to built a standalone version of the MC generator, 
{\tt all64} - to compile everything including default, ranlux-testing program
and seed-production and {\tt test64} - to compile everything and execute the test run scripts.

A simple way to compile the program is to issue 
{\tt make default64},
 being in the directory where the {\tt Makefile} is located.
This will produce {\tt ekhara.exe} (the main program executable) and copy it
into the sub-directory {\tt EXE}, together with the content of the
{\tt Env} sub-directory.
The latter contains the set of sample input files and 
histogram plotting scripts.
We provide a full set of necessary 
input files in the distribution package.
It is advised to execute {\tt ekhara.exe}
in the directory {\tt EXE},
where it is placed by default.
Every time one executes {\tt make default64}, the input files 
in  the directory {\tt EXE} are replaced with the sample ones from
the directory {\tt Env}.

EKHARA needs a random seed for operation.
Different random seeds can be obtained
by using the {\tt Makefile} target {\tt seed\_prod-ifort}.
It produces an executable program {\tt seed\_prod.exe},
which generates a set of random seeds.

\subsection{A test run description}
\label{sec:Testrun}

It is recommended to test the random 
number generator on a given machine,
before using EKHARA.
It is also important to check whether 
EKHARA can function properly on a given operational system
and that there are no critical bugs due to 
the compiler.
We provide a test run package for these purposes.

It is suggested to use the {\tt Makefile} target
{\tt test64}.
This will automatically prepare and execute
the following two test steps.

The first step of the test run is 
the random number generator control.
The source file {\tt testlxf.for} contains
the {\tt ranlux} test routines. The random numbers are the only part
 of the code used in double precision.

The second step is the verification
if the user-compiled EKHARA can reproduce
the set of results, created by a well-tested copy 
of EKHARA in various modes.
The test run environment contains directory 
{\tt test}  with pre-calculated data for the comparison,
the random seed and input files for each mode. 
The script {\tt test.sh} 
executes the user-compiled {\tt ekhara.exe} in all 
the control modes and compares the output with previously stored results.

Please read carefully the output of the test run execution
in your console and be sure there are no warnings and/or error
messages.

\subsection{A customisation of the source code by a user}
\label{sec:User}

We leave for a user an option to customise the generator to
her/his needs by editing the source code file 
{\tt ekhara-routines/routines-user.inc.for}.
Notice that we always use explicit declaration of identifiers and
the {\tt implicit none} statement is written down in each routine.

In the file {\tt ekhara-routines/routines-user.inc.for}
one can change 
\begin{itemize}
\item the data-card reading
(routines {\tt read\_card\_1pi}, {\tt read\_card\_2pi}  and {\tt read\_card\_chi}),
\item the form-factor formula
(routine {\tt piggFF}),
\item the events reporting (routines {\tt reportevent\_1pi},  {\tt reportevent\_1pi\_1ph} and {\tt reportevent\_chi}),
\item the histograming (routines {\tt histo\_event\_1pi}, {\tt histo\_event\_2pi},  {\tt histo\_event\_1pi\_0ph}
and {\tt histo\_event\_1pi\_1ph}),
\item additional kinematic cuts
(routines {\tt ExtraCuts\_1pi} and {\tt ExtraCuts\_2pi}).
\end{itemize}

\section{Conclusions}
\label{six}
In this paper we have presented the upgrades of the EKHARA Monte Carlo generator.
The main result being the radiative corrections to the reactions $e^+e^-\to e^+e^-P $,
  with a new algorithm of the phase
  space generation for the reaction $e^+e^-\to e^+e^-P \gamma$.
  Comparisons with GGRESRC generator are also shown. Big
  differences are observed between the radiative corrections
    calculated by the EKHARA generator,
   which uses NLO exact formulae and
   the GGRESRC generator based on the structure function approach.
 
\section{Acknowledgements}
\label{seven}
 We would like to thank Achim Denig and Christoph Redmer for discussions
  on experimental aspects of the studies and 
   Christoph Redmer for indicating a bug in the generated
   azimuthal angles distributions in earlier versions of the code.
   The work of Sergiy Ivashyn at the early stages of this research is also
   acknowledged.
  This work was supported 
 in part by
the Polish National Science Centre, grant number DEC-2012/07/B/ST2/03867
and
German Research Foundation DFG under
Contract No. Collaborative Research Center CRC-1044.





\bibliographystyle{elsarticle-num}
\bibliography{mybibfile}

\begin{thebibliography}{10}
\expandafter\ifx\csname url\endcsname\relax
  \def\url#1{\texttt{#1}}\fi
\expandafter\ifx\csname urlprefix\endcsname\relax\def\urlprefix{URL }\fi
\expandafter\ifx\csname href\endcsname\relax
  \def\href#1#2{#2} \def\path#1{#1}\fi

\bibitem{Actis:2010gg}
S.~Actis, et~al., {Quest for precision in hadronic cross sections at low
  energy: Monte Carlo tools vs. experimental data}, Eur. Phys. J. C66 (2010)
  585--686.
\newblock \href {http://arxiv.org/abs/0912.0749} {\path{arXiv:0912.0749}},
  \href {http://dx.doi.org/10.1140/epjc/s10052-010-1251-4}
  {\path{doi:10.1140/epjc/s10052-010-1251-4}}.

\bibitem{Bennett:2006fi}
G.~W. Bennett, et~al., {Final Report of the Muon E821 Anomalous Magnetic Moment
  Measurement at BNL}, Phys. Rev. D73 (2006) 072003.
\newblock \href {http://arxiv.org/abs/hep-ex/0602035}
  {\path{arXiv:hep-ex/0602035}}, \href
  {http://dx.doi.org/10.1103/PhysRevD.73.072003}
  {\path{doi:10.1103/PhysRevD.73.072003}}.

\bibitem{Keshavarzi:2018mgv}
A.~Keshavarzi, D.~Nomura, T.~Teubner, {The muon $g-2$ and $\alpha(M_Z^2)$: a
  new data-based analysis.}\href {http://arxiv.org/abs/1802.02995}
  {\path{arXiv:1802.02995}}.

\bibitem{Hagiwara:2017zod}
K.~Hagiwara, A.~Keshavarzi, A.~D. Martin, D.~Nomura, T.~Teubner, {g-2 of the
  muon: status report}, Nucl. Part. Phys. Proc. 287-288 (2017) 33--38.
\newblock \href {http://dx.doi.org/10.1016/j.nuclphysbps.2017.03.039}
  {\path{doi:10.1016/j.nuclphysbps.2017.03.039}}.

\bibitem{Jegerlehner:2017gek}
F.~Jegerlehner, {The Anomalous Magnetic Moment of the Muon}, Springer Tracts
  Mod. Phys. 274 (2017) pp.1--693.
\newblock \href {http://dx.doi.org/10.1007/978-3-319-63577-4}
  {\path{doi:10.1007/978-3-319-63577-4}}.

\bibitem{Davier:2017zfy}
M.~Davier, A.~Hoecker, B.~Malaescu, Z.~Zhang, {Reevaluation of the hadronic
  vacuum polarisation contributions to the Standard Model predictions of the
  muon $g-2$ and ${\alpha (m_Z^2)}$ using newest hadronic cross-section data},
  Eur. Phys. J. C77~(12) (2017) 827.
\newblock \href {http://arxiv.org/abs/1706.09436} {\path{arXiv:1706.09436}},
  \href {http://dx.doi.org/10.1140/epjc/s10052-017-5161-6}
  {\path{doi:10.1140/epjc/s10052-017-5161-6}}.

\bibitem{Gioiosa:2017oda}
A.~Gioiosa, A.~Anastasi, {Construction and Commissioning of the beam delivery,
  storage ring and g-2 detectors at FNAL}, PoS EPS-HEP2017 (2017) 664.
\newblock \href {http://dx.doi.org/10.22323/1.314.0664}
  {\path{doi:10.22323/1.314.0664}}.

\bibitem{TI}
\href{https://indico.fnal.gov/event/13795/}{g-2 Theory Initiative}.
\newline\urlprefix\url{https://indico.fnal.gov/event/13795/}

\bibitem{Aubert:2009mc}
B.~Aubert, et~al., {Measurement of the gamma gamma* ---> pi0 transition form
  factor}, Phys. Rev. D80 (2009) 052002.
\newblock \href {http://arxiv.org/abs/0905.4778} {\path{arXiv:0905.4778}},
  \href {http://dx.doi.org/10.1103/PhysRevD.80.052002}
  {\path{doi:10.1103/PhysRevD.80.052002}}.

\bibitem{BABAR:2011ad}
P.~del Amo~Sanchez, et~al., {Measurement of the $\gamma \gamma^* --> \eta$ and
  $\gamma \gamma* --> \eta'$ transition form factors}, Phys. Rev. D84 (2011)
  052001.
\newblock \href {http://arxiv.org/abs/1101.1142} {\path{arXiv:1101.1142}},
  \href {http://dx.doi.org/10.1103/PhysRevD.84.052001}
  {\path{doi:10.1103/PhysRevD.84.052001}}.

\bibitem{Uehara:2012ag}
S.~Uehara, et~al., {Measurement of $\gamma \gamma^* \to \pi^0$ transition form
  factor at Belle}, Phys. Rev. D86 (2012) 092007.
\newblock \href {http://arxiv.org/abs/1205.3249} {\path{arXiv:1205.3249}},
  \href {http://dx.doi.org/10.1103/PhysRevD.86.092007}
  {\path{doi:10.1103/PhysRevD.86.092007}}.

\bibitem{TREBSBST}
S.~Uehara, TREBSBST, , KEK Report, No. 96-11 (1996).

\bibitem{Druzhinin:2014sba}
V.~P. Druzhinin, L.~V. Kardapoltsev, V.~A. Tayursky, {GGRESRC: A Monte Carlo
  generator for the two-photon process $e^ + e^- \to e^ + e^- R(J^{PC}=0^{-+})$
  in the single-tag mode}, Comput. Phys. Commun. 185 (2014) 236--243.
\newblock \href {http://dx.doi.org/10.1016/j.cpc.2013.07.017}
  {\path{doi:10.1016/j.cpc.2013.07.017}}.

\bibitem{Rodrigo:2001kf}
G.~Rodrigo, H.~Czyz, J.~H. Kuhn, M.~Szopa, {Radiative return at NLO and the
  measurement of the hadronic cross-section in electron positron annihilation},
  Eur. Phys. J. C24 (2002) 71--82.
\newblock \href {http://arxiv.org/abs/hep-ph/0112184}
  {\path{arXiv:hep-ph/0112184}}, \href
  {http://dx.doi.org/10.1007/s100520200912} {\path{doi:10.1007/s100520200912}}.

\bibitem{Czyz:2012nq}
H.~Czyz, S.~Ivashyn, A.~Korchin, O.~Shekhovtsova, {Two-photon form factors of
  the $\pi^0$, $\eta$ and $\eta'$ mesons in the chiral theory with resonances},
  Phys. Rev. D85 (2012) 094010.
\newblock \href {http://arxiv.org/abs/1202.1171} {\path{arXiv:1202.1171}},
  \href {http://dx.doi.org/10.1103/PhysRevD.85.094010}
  {\path{doi:10.1103/PhysRevD.85.094010}}.

\bibitem{Czyz:2017veo}
H.~Czyz, P.~Kisza, S.~Tracz, {Modeling interactions of photons with
  pseudoscalar and vector mesons}, Phys. Rev. D97~(1) (2018) 016006.
\newblock \href {http://arxiv.org/abs/1711.00820} {\path{arXiv:1711.00820}},
  \href {http://dx.doi.org/10.1103/PhysRevD.97.016006}
  {\path{doi:10.1103/PhysRevD.97.016006}}.

\bibitem{Czyz:2010sp}
H.~Czyz, S.~Ivashyn, {EKHARA: A Monte Carlo generator for $e^+ e^- \to e^+ e^-
  \pi^0$ and $e^+ e^- \to e^+ e^- \pi^+ \pi^-$ processes}, Comput. Phys.
  Commun. 182 (2011) 1338--1349.
\newblock \href {http://arxiv.org/abs/1009.1881} {\path{arXiv:1009.1881}},
  \href {http://dx.doi.org/10.1016/j.cpc.2011.01.029}
  {\path{doi:10.1016/j.cpc.2011.01.029}}.

\bibitem{Schuler:1997ex}
G.~A. Schuler, {Two photon physics with GALUGA 2.0}, Comput. Phys. Commun. 108
  (1998) 279--303.
\newblock \href {http://arxiv.org/abs/hep-ph/9710506}
  {\path{arXiv:hep-ph/9710506}}, \href
  {http://dx.doi.org/10.1016/S0010-4655(97)00127-6}
  {\path{doi:10.1016/S0010-4655(97)00127-6}}.

\bibitem{Czyz:2016xvc}
H.~Czyz, J.~H. K{\"u}hn, S.~Tracz, {$\chi_{c1}$ and $\chi_{c2}$ production at
  $e^+e^-$ colliders}, Phys. Rev. D94~(3) (2016) 034033.
\newblock \href {http://arxiv.org/abs/1605.06803} {\path{arXiv:1605.06803}},
  \href {http://dx.doi.org/10.1103/PhysRevD.94.034033}
  {\path{doi:10.1103/PhysRevD.94.034033}}.

\bibitem{Czyz:2016lwq}
H.~Czyz, P.~Kisza, {Testing $\chi_c$ properties at BELLE II}, Phys. Lett. B771
  (2017) 487--491.
\newblock \href {http://arxiv.org/abs/1612.07509} {\path{arXiv:1612.07509}},
  \href {http://dx.doi.org/10.1016/j.physletb.2017.05.091}
  {\path{doi:10.1016/j.physletb.2017.05.091}}.

\bibitem{Barbieri:1972as}
R.~Barbieri, J.~A. Mignaco, E.~Remiddi, {Electron form-factors up to fourth
  order. 1.}, Nuovo Cim. A11 (1972) 824--864.
\newblock \href {http://dx.doi.org/10.1007/BF02728545}
  {\path{doi:10.1007/BF02728545}}.

\bibitem{Bonciani:2003ai}
R.~Bonciani, P.~Mastrolia, E.~Remiddi, {QED vertex form-factors at two loops},
  Nucl. Phys. B676 (2004) 399--452.
\newblock \href {http://arxiv.org/abs/hep-ph/0307295}
  {\path{arXiv:hep-ph/0307295}}, \href
  {http://dx.doi.org/10.1016/j.nuclphysb.2003.10.031}
  {\path{doi:10.1016/j.nuclphysb.2003.10.031}}.

\bibitem{vanNeerven:1984ak}
W.~L. van Neerven, J.~A.~M. Vermaseren, {The Role of the Five Point Function in
  Radiative Corrections to Two Photon Physics}, Phys. Lett. 142B (1984) 80.
\newblock \href {http://dx.doi.org/10.1016/0370-2693(84)91140-7}
  {\path{doi:10.1016/0370-2693(84)91140-7}}.

\bibitem{CKS}
H.~Czyz, K.~Kampf, P.~Sanchez-Puertas, in preparation.

\bibitem{alphaQED}
F.~Jegerlehner,
  \href{http://www-com.physik.hu-berlin.de/~fjeger/software.html}{alphaQEDc17}
  (October 2017).
\newline\urlprefix\url{http://www-com.physik.hu-berlin.de/~fjeger/software.html}

\bibitem{Jegerlehner:2017zsb}
F.~Jegerlehner, {Variations on Photon Vacuum Polarization}\href
  {http://arxiv.org/abs/1711.06089} {\path{arXiv:1711.06089}}.

\bibitem{Jegerlehner:2011mw}
F.~Jegerlehner, {Electroweak effective couplings for future precision
  experiments}, Nuovo Cim. C034S1 (2011) 31--40.
\newblock \href {http://arxiv.org/abs/1107.4683} {\path{arXiv:1107.4683}},
  \href {http://dx.doi.org/10.1393/ncc/i2011-11011-0}
  {\path{doi:10.1393/ncc/i2011-11011-0}}.

\bibitem{book}
E.~Byckling, K.~Kajantie, Particle Kinematics, 1st Edition, John Wiley and Sons
  Ltd, 1973.

\bibitem{Czyz:2006dm}
H.~Czyz, E.~Nowak-Kubat, {The Reaction $e^+ e^- \to e^+ e^- \pi^+ \pi^-$ and
  the pion form-factor measurements via the radiative return method}, Phys.
  Lett. B634 (2006) 493--497.
\newblock \href {http://arxiv.org/abs/hep-ph/0601169}
  {\path{arXiv:hep-ph/0601169}}, \href
  {http://dx.doi.org/10.1016/j.physletb.2006.02.024}
  {\path{doi:10.1016/j.physletb.2006.02.024}}.

\bibitem{Kuipers:2012rf}
J.~Kuipers, T.~Ueda, J.~A.~M. Vermaseren, J.~Vollinga, {FORM version 4.0},
  Comput. Phys. Commun. 184 (2013) 1453--1467.
\newblock \href {http://arxiv.org/abs/1203.6543} {\path{arXiv:1203.6543}},
  \href {http://dx.doi.org/10.1016/j.cpc.2012.12.028}
  {\path{doi:10.1016/j.cpc.2012.12.028}}.

\end{thebibliography}


\begin{thebibliography}{0}
\bibitem{1} H. Czy\.z, S. Ivashyn, Comput. Phys. Commun. 182 (2011) 1338.  
\bibitem{2} H. Czy\.z, E. Nowak-Kubat, Phys. Lett. B634 (2006) 493.
\bibitem{3} H. Czy\.z, S. Ivashyn, A. Korchin, O. Shekhovtsova, Phys. Rev. D85 (2012) 094010         
\bibitem{4} H. Czy\.z, J.H. K\"uhn, Sz. Tracz, Phys.Rev. D94 (2016) 034033
\bibitem{5} H. Czy\.z, P. Kisza, Phys. Lett. B771 (2017) 487
\bibitem{6} H. Czy\.z, P. Kisza,  Sz. Tracz, Phys. Rev. D97 (2018) 016006
\end{thebibliography}







\end{document}